# Are there alternatives to our present theories of physical reality?

Peter Rowlands

Department of Physics, University of Liverpool, Oliver Lodge Laboratory, Oxford Street, Liverpool, L69 7ZE, UK. e-mail: p.rowlands@liverpool.ac.uk

**Abstract**

Our notions of what is physically 'real' have long been based on the idea that the real is what is immediately apprehended, that is the local or observable, the physically tangible, though there has always been an alternative philosophical notion that the 'real' is some kind of ontological structure beyond immediate apprehension, and so inaccessible through physics. However, quantum mechanics, with its intrinsic nonlocal correlations, has seemingly left us with a dilemma by showing that fundamental physical theories cannot be both real and local. Reality cannot be reconstructed as a deterministic projection from physical observations. Many people think that the problem lies with quantum mechanics, but, in fact, it is more likely to be a result of unrealistic expectations. We have assumed that fundamental physics ought to be compatible with normal (macroscopic) experience. If, however, we go beyond our current high-level physical theories to the basic elements from which they are constructed, we see that a pattern emerges that gives us a very different and much more coherent understanding of what is meant by physical 'reality'.

## 1 Introduction

Many people would say that current physical theories give us problems in defining the meaning of physical reality. However, it may be that we are effectively looking through the wrong end of a telescope. We are treating our sophisticated 'high level' theories as the fundamental language, rather than looking at the more basic elements from which they are constructed. I am going to do something very unusual, though I don't fully understand why it should be so. I am going to look at physics in a way that no one seems to realise is of real interest. But it seems to me that the answer to many of the most fundamental questions in physics lies at a deeper level than we have so far penetrated.

The point where we want to start is not any of the high-level theories and concepts, like the Standard Model, quantum mechanics, general relativity, etc. Certainly we want to get to such theories, but rather as *emergent* aspects of something deeper and simpler, than as the starting points of our investigation. Why haven't we done this already? I simply don't know. It is not obvious, of course, that it can be done, but it seems obvious to me that we should try for it. Our present theories of reality give us conceptual problems because we don't truly understand the significance of the foundations on which they are built. But, in addition, previous experience tells us that getting to a more fundamental level is also the key to considerable technical as well as conceptual advances.

An elementary, but powerful, technique used in numerous aspects of physics – dimensional analysis – suggests that some physical concepts are more fundamental than others. But these include concepts such as space and time which no one has been able to penetrate. Can we get any further? Not, it would seem, mathematically. But there is one other option, and it is one already used to great effect in physics – symmetry. Symmetry allows us to reduce assumptions, and to cut out levels of explanation. Though we don't totally understand why it is effective, it is certainly a component of all the most fundamental theories. Does it reach the most fundamental level of all? Can we do a 'dimensional analysis of symmetry', rather than of algebra?

In principle, physics (and not just quantum physics) has always been structured on an opposition between an abstract mathematical system and a process of physical measurement – an idea related to the philosophical opposition between ontology and epistemology. A study of the symmetries relating to the most basic physical concepts might give us an insight into the way in which this opposition arises.

Hardly anyone would deny that space and time are among the most fundamental of all physical notions. They are also the only means we have of observing change and variation in nature – the basis of all science. Space, in addition, is the only quantity that is ever actually measured. Even time 'measurements' really only amount to observations of elements of space. So, what else is really fundamental in nature? There can only be one answer – the sources of the four fundamental interactions, the causative agents for the variations which we observe in space and time. For reasons that will become apparent, I will refer to these as 'mass' and the three 'components' (electric, strong and weak) of a composite parameter 'charge' (as in 'charge conjugation').

Let us assume: (1) that enough iterations have been done to establish that this is the correct starting point, and that the assumptions will ultimately be justified by the results; (2) that our sophisticated 'high level' theories will emerge as constructed from 'packages' composed from such elements; (3) that symmetry-breaking is an aspect of the packaging and not of the fundamental nature of the constituents. Here, mass is regarded as the source of the gravitational field, i.e. mass-energy, not rest mass (which is, of course, never observed anyway). Charge is a generic term for the sources of the electric, strong and weak interactions, and so behaves as a kind of 3-dimensional parameter. The perfect symmetry between the electric, strong and weak charges is broken at normal energies, but it is believed that, under ideal conditions (grand unification), all 3 charge terms would be exactly alike. The simplest possible ideas, then, are space, time, mass and charge.

Space, as we have already said, has a unique property. It is the only parameter that can be measured. Every other so-called measurement becomes a matter of observing a pointer moving over a scale or equivalent. Any object whatsoever sets up a measurement of space. Time 'measurement', however, requires special conditions in which we count repetitions of the same interval. Measurability is not a universal aspect of nature, nor is anything else. That is why we also need time, mass and charge.

**2 Conserved and Nonconserved: Mass / Charge and Space / Time**

All the fundamental laws of physics are concerned in some way with conservation and nonconservation. Now, nonconservation is not simply the absence of conservation but a property with equally definite characteristics. Conserved quantities are conserved *locally*; their units have individual identities which remain after space and time variations. Nonconserved quantities have *no identity*. One unit of the quantity is as good as any other. So, we have

>    the translation symmetry of time
>    the translation symmetry of space
>    the rotation symmetry of space

Conserved quantities, by contrast, are translation and rotation asymmetric. Each unit is unique. One cannot be replaced by another. So, we have

>    the translation *a*symmetry of mass
>    the translation *a*symmetry of charge
>    the rotation *a*symmetry of charge

The last is especially important. The three types of charge do not rotate into each other. The 3 types of charge are separately conserved. The baryon and lepton conservation laws illustrate this. Baryons are the only fermions with strong charges, so cannot decay into fermions without them. Leptons are the only fermions with weak but no strong charges, so cannot be created from baryons.

Another key property of nonconserved quantities is *gauge invariance*. So, for example, field terms remain unchanged under arbitrary changes in the vector and scalar potentials, or phase changes in the quantum mechanical wavefunction, due to translations (or rotations) in the space and time coordinates. So a system will remain conservative under arbitrary changes in *the coordinates* which don't produce changes in the values of conserved quantities such as charge, energy, momentum and angular momentum. Significantly, gauge invariance is *local*, like the conservation laws.

In general, physics structures itself in terms of *differential equations* which ensure that the conserved quantities – mass and charge, and others derived from them, such as energy, momentum and angular momentum – remain unchanged while the nonconserved or variable quantities vary absolutely. This means that the nonconserved or variable quantities are expressed in physics equations as *differentials*, *dx*, *dt*, directly expressing this variation.

The idea that 'God plays dice' in the quantum state should no longer trouble us if we accept the logic of defining space and time as nonconserved quantities. This means that they are not fixed and should be subject to absolute variation. It is only the fact that conservation principles should hold at the same time that restricts the range of variation when systems interact with each other. When the interactions are on a massive scale, we can even make a classical 'measurement'. This is how I tend to see the so-called 'collapse of the wavefunction'. I don't tend to think that 'measuring apparatus' (as in the Copenhagen interpretation) needs to be an intrinsic aspect of quantum mechanics. We simply apply conservation conditions (via potentials) to restrict the degree of variability that would otherwise exist (decoherence).

According to that well-known mathematical result, Noether's theorem, to every variational property there is a conserved quantity. So

> translation symmetry of time ≡ conservation of energy
> translation symmetry of space ≡ conservation of momentum
> rotation symmetry of space ≡ conservation of angular momentum

Noether's theorem is a natural consequence of defining conservation and nonconservation properties symmetrically; and nonconservation of time ≡ conservation of mass (energy) is one of several examples of how we might expect it to operate.

## 3 Real and Imaginary: Mass / Space and Time / Charge

The real / imaginary distinction is between parameters whose squared values are positive (norm 1) and those whose squared values are negative (norm –1). It is an intrinsically mathematical, rather than a physical distinction, though it does have physical consequences. Special relativity combines space and time in a *4-vector*, with 3 real parts (space) and one imaginary part (time). Pythagoras' theorem in 4-D

$$r^2 = x^2 + y^2 + z^2 - c^2t^2 = x^2 + y^2 + z^2 + i^2c^2t^2$$

leads naturally to a 4-vector representation for space and time:

$$r = \mathbf{i}x + \mathbf{j}y + \mathbf{k}z + ict.$$

This is sometimes described as a mere mathematical 'trick', but we have in that case to explain why the 'trick' works. And the real / imaginary distinction is not restricted to space and time. The long-established but totally unexplained fact that identical masses attract; identical charges repel has a natural explanation if we suppose that masses are real, and charges imaginary.

But there are three charges: electric, strong and weak. They are all alike in being mutually repelling when identical. So, we have to assume that there must be some way of having a 3-D imaginary quantity, and, of course, there is, as Hamilton discovered in 1843. This is the *quaternion* system, with 3 imaginary parts and one real, and it has the additional value of being unique. The imaginary part can *only* be 3-D – unless we want to break algebraic associativity in extending to octonions. The quaternions are cyclic and anticommutative, properties that are also physically significant:

$$i^2 = j^2 = k^2 = ijk = -1$$
$$ij = -ji = k$$
$$jk = -kj = i$$
$$ki = -ik = j$$

Hamilton discovered quaternions when he was trying to extend the complex number system to explain 3-D space. He believed the imaginary part represented space and the real part time. He thought that he had discovered a key to the universe – and the prevailing opinion ever since has been that this was a massive delusion. 'Never has a great mathematician been proved more hopelessly wrong', wrote E. T. Bell in his popular work, *Men of Mathematics*.[1] For the best part of a century, virtually everyone (except thousands of computer software engineers!) has automatically agreed with Bell. Quaternions may be clever, but they don't have much physical significance, except as leading to things, like vectors, which have. But it was the mediocre historian of science who was 'hopelessly wrong', not the great mathematician. Space and time become a *4-vector* with three real parts and one imaginary, by symmetry with the mass and charge quaternion, with three imaginary parts and one real.

| space | time | charge | mass |
|---|---|---|---|
| **i**x **j**y **k**z | *i*t | *i*s *j*e *k*w | 1m |

And there is a bonus. If we apply quaternionic multiplication rules to the space vector, so that it becomes a multivariate Clifford algebra, as Hestenes did in 1966,[2] and as *we* require for absolute symmetry, we also automatically incorporate the otherwise strange property of *spin*. Multivariate vectors, unlike standard ones, have a full product, defined by

$$\mathbf{ab} = \mathbf{a}.\mathbf{b} + i\mathbf{a} \times \mathbf{b}$$

When multivariate vectors are introduced into the space / momentum terms of relativistic quantum mechanics, in the disguised form of Pauli matrices, the additional cross product term *i***a** × **b** turns out to be the one which leads to fermionic half-integral spin. Space and time are simply quaternions multiplied by *i*, and spin is simply a topological property of space (as Dirac knew), and not quantum or relativistic in origin.

There are even further advantages in the real / imaginary description of physical quantities. For example, only quantities in which the time is *squared* are significant in physics – acceleration and force. Time 'measurement' always requires force and acceleration. There is no one-way speed of light. Also, imaginary quantities are algebraically dual (unlike real ones): that is, + solutions only exist if there are also − ones. This means that *all* charges (not just electric) have to have solutions for *both* signs. We have to have antiparticles or antistates, both with positive mass. Again, we have two ways of detecting *real* mass – directly, though inertia, and via the squared quantity (gravity) – but only one way of detecting imaginary charge, via the squared quantity (electric force, etc.).

**4 Divisible and Indivisible: Space / Charge and Mass / Time**

Mass, in the sense of mass-energy, is a continuum. It is present at all points in space. There are the Higgs field (246 GeV) or vacuum, the 2.7 K microwave background radiation, the zero-point energy, even ordinary fields. The continuity of mass is the precise reason why it can never be negative (or, more strictly, change sign). There is no zero or crossover point. Charge, however, has always been recognized as being discrete and being delivered in precise units.

Similarly, space and time are fundamentally different, and not just mathematically, as real and imaginary quantities. The root cause is that time is continuous and space is not. Time's continuity has many consequences. It means that time is irreversible. To reverse time, we would have to create a discontinuity, a zero-point, and it would no longer be continuous. Time also is not an observable in quantum mechanics, because observables must be discrete. And it is always treated as the independent variable; we write $dx/dt$, not $dt/dx$.

The absolute continuity of time is important in the explanation of the paradox of Zeno in which Achilles never catches the tortoise, however fast he runs, if he gives it a start, and the same is true of other paradoxes of a similar nature. Various authors have seen that the problem lies in the assumption that one can divide time into observational units like space. Whitrow, for example,[3] writes that: 'One can, therefore, conclude that the idea of the infinite divisibility of time must be rejected, or ... one must recognize that it is ... a logical fiction.' Motion is 'impossible if time (and, correlatively, space) is divisible ad infinitum'. And Coveney and Highfield[4] propose that: 'Either one can seek to deny the notion of 'becoming', in which case time assumes essentially space-like properties; or one must reject the assumption that time, like space, is infinitely divisible into ever smaller portions.' Perhaps because of the many historical efforts to link space and time in a more than mathematical sense, such authors seem to be reluctant to draw the logical conclusion that the paradox, like many others, really is a result of making things that are fundamentally unlike have the same properties. Space is 'infinitely divisible into ever smaller portions'; time is not divisible at all. What we call 'divisions of time' are not observed through time at all.

Again, all normal physical equations are time-reversible, but *time* is not. We know this from the second law of thermodynamics. This has been regarded as a paradox, the so-called 'reversibility paradox', but an investigation into the fundamental structures provides a simple explanation. Physical equations are time-reversible mathematically, because time is an imaginary parameter with equal + and − solutions; and, of course, the action of physical forces always involves time squared, so + or − makes no difference. However, time itself, as a continuum can never be reversed. There is thus no paradox.

Space has to be discrete, because it could not otherwise be observed. However, its discreteness is different from that of charge because it is a nonconserved quantity and so has no fixed units. This means that its discreteness must be endlessly reconstructed. In other words, it is *infinitely* divisible. It is the absolute continuity of time which denies it this property. Infinite divisibility is the absolute opposite of continuity.

But, we may ask, is space not represented mathematically as a real number line? The answer is that it is (because of nonconservation), but real numbers are not necessarily absolutely continuous, as supposed by Cantor. They can be defined that way, but there is an equally valid way of defining them to be algorithmically countable, as established by Skolem in 1934, and as applied subsequently by Robinson to calculus.[5] There are two systems of algebra, two of geometry and two of calculus, which depend on two different, equally valid definitions of the real numbers. They are called Standard and Nonstandard Analysis, and there is a perfect duality between them.

| Standard | Nonstandard |
|---|---|
| Continuous | Discrete |
| Noncountable reals | Countable reals |
| Cantor | Robinson |
| Limits | Infinitesimals |
| Time | Space |

There were two ways of differentiating, known from the seventeenth century, based respectively on the properties of time and space. The theory of infinitesimals, as developed by Robinson and others, is now a fully-accepted part of mathematics, and considered just as rigorous as the theory of limits. Significantly, only the time-based method of analysis (limits) solves Zeno.

The *duality* we have seen at the foundational level in mathematics applies in exactly the same way in physics. When we combine space and time in a 4-vector, we are really doing something that is mathematically possible, but physically impossible. So we either make time spacelike – the discrete solution – or space timelike – the continuous solution. This is the origin of wave-particle duality. Hermann Minkowski famously said: 'From now on, space by itself, and time by itself, are destined to sink into shadows, and only a kind of union of both to retain an independent existence'.[6] However, this statement cannot be considered valid from a physical point of view. The many physical manifestations of wave-particle duality clearly deny it. The choice between discrete and continuous options occurs in practically every area of physics, and no method has so far been found to validate one at the expense of the other.

|   | Discrete options | Continuous options |
|---|---|---|
|   | particles | waves |
|   | relativity | Lorentzian aether |
|   | Heisenberg | Schrödinger |
|   | amplitude | phase |
|   | quantum electrodynamics | stochastic electrodynamics |
|   | $h/2\pi$ | $h/4\pi$ |
|   | potential energy | kinetic energy |
|   | charge-like | mass-like |
|   | space-like | time-like |
|   | momentum-related | energy-related |
|   | spin 1 exchange | spin ½ exchange |

Nature is neither totally continuous nor totally discrete. Attempts have been made many times to claim that physics would be better if made totally discrete, but continuity always forces its way in (e.g. through the second law of thermodynamics). If we make the system discrete (as Heisenberg) did, continuity appears in the measurement (Heisenberg uncertainty). If we make the system continuous (like Schrödinger), then discreteness appears in the measurement (collapse of the wavefunction). In the diagram below, the terms in bold type are the physically 'incorrect' options in the system that must be reversed in the process of measurement.

The discrete option: Heisenberg's quantum mechanics

|   | The System |   | Measurement |   |
|---|---|---|---|---|
| discrete | space | real | not changed |   |
|   | charge | particles | by measurement |   |
|   | momentum |   |   |   |
|   | angular momentum |   |   |   |
|   | **time** | **virtual** | restores | introduces |
|   | **mass** | **vacuum** | continuity | nonlocalised |
|   | **energy** |   | of these | vacuum |

The continuous option: Schrödinger's wave mechanics

|   | The System |   | Measurement |   |
|---|---|---|---|---|
| **continuous** | **space** | **virtual** | restores | introduces |
|   | **charge** | **particles** | discreteness | localised |
|   | **momentum** |   | of these | particles |
|   | **angular momentum** |   |   |   |
|   | time | real | not changed |   |
|   | mass | vacuum | by measurement |   |
|   | energy |   |   |   |

Remarkably, discrete quantities appear to be (3-)dimensional while continuous quantities are non-dimensional. It is easy to see why continuous quantities cannot have dimensions – dimensionality requires an origin, a zero or crossover point, which is incompatible with continuity. But, why are discrete quantities 3-D? We can see this in a roundabout way, by noticing that a quantity with only 1-D could not be measured, because the crossover points to another dimension are needed to do the scaling. So, a line is not actually a 1-D structure, but a 1-D structure that can only exist in a 2-D world. Then the 3-D extension is required for symmetry with quaternions.

However, there is a direct argument which is much more profound and takes us to the very deepest foundations of both mathematics and physics. Essentially, discreteness in physics comes only from anticommutativity (a principle which is, of course, significant for quantum mechanics). If **a** is anticommutative with **b**, then (if we ignore scalar multiplication) it cannot be anticommutative with anything else except **ab**. **a**, **b** and **ab** form a closed (discrete) set. The full significance of this is explained in the first chapter of *Zero to Infinity*,[7] which sets out to derive the symmetries between the fundamental parameters from even more foundational arguments.

## 5 A group of order 4

The results so far presented suggest that the four fundamental parameters can be organized around 3 fundamental dualities, with a different pairing for each:

> Conserved / Nonconserved
> Real (norm 1) / Imaginary (norm –1)
> Commutative / Anticommutative

In fact, everywhere in physics where the factor 2 or ½ appears in a fundamental context, it emerges because of one of these dualities, and it is often possible to switch the explanation from one duality to another.

An extreme case is electron spin or the magnetic moment associated with it, where the different explanations include all of the dualities. Deriving it via the Dirac equation, which essentially means the anticommutativity of the momentum operator, we are using the commutative / anticommutative duality. Deriving it via the Thomas precession (i.e. relativity) introduces the real / imaginary duality. Alternatively we can invoke an extremely simple explanation of the magnetic effect using the distinction between kinetic and potential energy equations for changing and static conditions, which effectively uses the conserved / nonconserved duality. Another example is spontaneous emission where the coefficient is twice that for stimulated emission either because of radiation reaction (the conserved / nonconserved duality) or relativity (the real / imaginary duality). Ultimately, the explanation is duality itself, rather than any particular form of it.

The properties of the parameters show a symmetry which can be conveniently arranged in a symmetric structure:

| | | | |
|---|---|---|---|
| **mass** | conserved | real | commutative |
| **time** | nonconserved | imaginary | commutative |
| **charge** | conserved | imaginary | anticommutative |
| **space** | nonconserved | real | anticommutative |

Here, the properties real and imaginary are alternatively described as norm 1 and norm −1; while the properties commutative and anticommutative are alternatively described as nondimensional and dimensional or continuous and discrete. As we have shown, the symmetric options are exact opposites, and so can be conveniently described by algebraic symbols:

| | | | |
|---|---|---|---|
| **mass** | $x$ | $y$ | $z$ |
| **time** | $-x$ | $-y$ | $z$ |
| **charge** | $x$ | $-y$ | $-z$ |
| **space** | $-x$ | $y$ | $-z$ |

In algebraic terms, this is a conceptual zero (though the actual signs and symbols are of course arbitrary). It is also a finite noncyclic group of order 4 (D2, Klein-4), in which element is its own inverse. We can generate group multiplication rules of the form:

$$x * x = -x * -x = x$$
$$x * -x = -x * x = -x$$
$$x * y = y * -x = 0$$

and similarly for $y$ and $z$, to establish a group multiplication table of the form:

| * | mass | charge | time | space |
|---|---|---|---|---|
| mass | mass | charge | time | space |
| charge | time | mass | space | charge |
| time | charge | space | mass | time |
| space | space | time | charge | mass |

Here, we have privileged mass as the identity element, but we could equally well have privileged charge, time or space. Many representations are possible. For example, we can use the H4 algebra (which is effectively equivalent to quaternions without + and − signs), where mass, charge, time, and space, are respectively represented by the units 1, **i**, **j**, **k**.

| * | 1 | **i** | **j** | **k** |
|---|---|---|---|---|
| **1** | 1 | i | j | k |
| **i** | j | 1 | k | i |
| **j** | i | k | 1 | j |
| **k** | k | j | i | 1 |

We can also postulate a dual group, defined by the symbols:

$$x \quad -y \quad z$$
$$-x \quad y \quad z$$
$$x \quad y \quad -z$$
$$-x \quad -y \quad -z$$

representing dual elements with the properties:

| | | | |
|---|---|---|---|
| **mass dual** | conserved | imaginary | commutative |
| **time dual** | nonconserved | real | commutative |
| **charge dual** | conserved | real | anticommutative |
| **space dual** | nonconserved | imaginary | anticommutative |

The symmetry may be assumed to be absolutely exact – no exception to this rule has ever been found. And this condition can be used to put constraints on physics to derive laws and states of matter. We can also develop a number of representations, which not only show the absoluteness of the symmetry, but also the centrality to the whole concept of the idea of 3-dimensionality.

**6 Representations of the group**

A perfect symmetry between 4 parameters means that only the properties of one parameter need be assumed. The others then emerge automatically like kaleidoscopic images. It is, in principle, arbitrary which parameter we assume to begin with, as the following visual representations will show. The representations also suggest that 3-dimensionality is a fundamental component of the symmetry. In the first, we represent the four parameters, space, time, mass and charge, by concentric circles, arbitrarily choosing the identity element as occupying the centre circle. Each circle is then divided into three sectors, with the properties / antiproperties identified by different primary / secondary colours. A reversal of these could be taken as representing the dual group. The totality in each sector always adds to zero (represented by white).

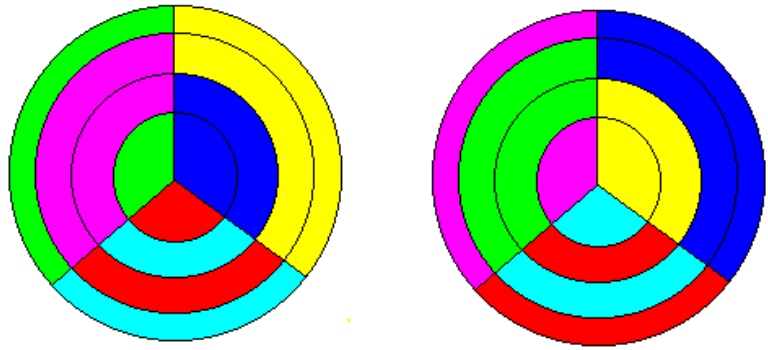

   The second representation shows *x*, *y*, *z* directions as emerging from the centre of a cube, and then plots the + and – values of *x*, *y*, *z* associated with each parameter in the algebraic representation, so that the four parameters become four solid lines drawn from this origin to four corners of the cube. The dotted lines drawn to the other four corners then represent the members of the dual group.

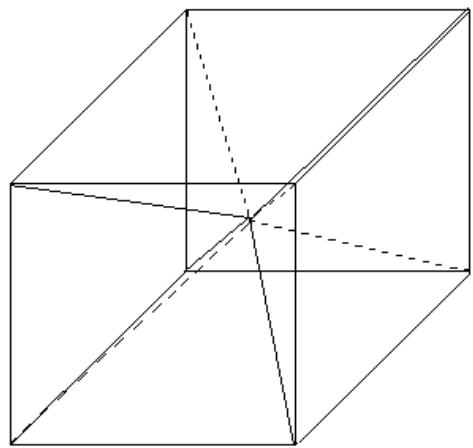

   A third representation situates the parameters at the vertices of a regular tetrahedron, with the six edges 'coloured' to represent the properties (say the primary colours, R, G, B) and antiproperties (say the secondary colours, M, C, Y). The four faces then become the members of the dual group. Of course, these representations could be reversed, so that the faces become space, time, mass and charges and the vertices the members of the dual group, or that the properties are represented by M, C, Y and the antiproperties by R, G, B.

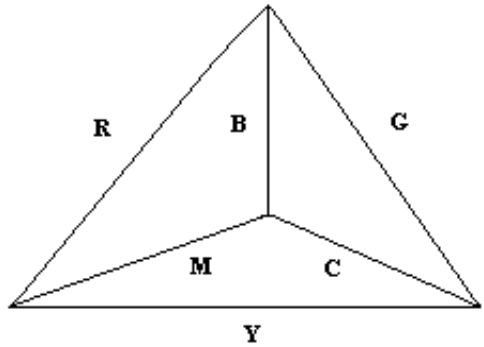

**7 Some consequences of the symmetry**

It is convenient to summarise a few of the many results we can obtain purely from the symmetry:

>God does play dice!
>Irreversibility of time
>Dual status of real numbers
>Explanation of the reversibility paradox
>Explanation of wave-particle duality

However, not only does the symmetry produce such conceptual explanations, it also leads to new *mathematical* results; e.g. not only does it lead to Noether's theorem, but also to two new extensions of it. We can extend Noether's theorem, purely by symmetry:

| conservation of | conservation of | symmetry of |
|---|---|---|
| energy | mass | time translation |
| momentum | magnitude of charge | space translation |
| angular momentum | type of charge | space rotation |

The first of these already has a partial realisation. In 1927 Fritz London showed that conservation of electric charge was identical to invariance under transformations of electrostatic potential by a constant representing changes of phase – of the kind involved in conservation of momentum.

The second result seems totally bizarre. How can the conservation of angular momentum be the same thing as the conservation of *type* of charge? In fact angular momentum conservation is *three separate conservation laws* – of magnitude; of direction; and of handedness. And these are precisely those involved in the $U(1)$, $SU(3)$ and $SU(2)$ symmetries involved with the electric, strong and weak charges. In principle, the conservation laws of magnitude, of direction, and of handedness, say, respectively, that the spherical symmetry of 3-dimensional space is preserved by a rotating system whatever the length of the radius vector;
whatever system of axes we choose; and
whether we choose to rotate the system left- or right-handed and these considerations are totally independent of each other.

The whole point of developing the symmetry argument about space, time, mass and charge becomes apparent when we propose that it is both exact and exclusive. So, according to this argument, not only can it not be broken. It must also be the sole source of information from which physics at higher levels can be constructed. However, we have also implied that such 'higher level' information cannot be created without taking the whole structure as a package. Immediately, we see why we have such difficult in physics in separating ontology and epistemology. The conserved quantities represent an ontology, while the nonconserved ones define the meaning of epistemology, and they cannot be separated.

Nature or 'reality' cannot be defined as either ontological or epistemological – the source of a traditional philosophical dilemma on the meaning of 'reality'. Also, we cannot assume everything can be measured or observed. If we, say, treat space and time as an epistemological 'probe' for understanding nature, the system provides us with sources which act as an ontological 'response'. Physics is specifically constructed to avoid creating any specific concept of reality. Nevertheless, circumscribed beings like ourselves cannot avoid thinking in 'realistic' terms. So we have created a system in which apparent reality in one aspect is countered by total nonreality in another. So, at the most fundamental level, physics is described by an abstract system, whose relationship to the original concept of measurement is only ever indirect, though it must always be present. Measurement is a component of the system, but it cannot describe it completely.

**8 The nilpotent structure**

The system only works when treated as a 'package', which is specifically put together to prevent arbitrary characterization of nature. So we should ask whether such a 'packaging' helps us to a view of nature that works well with such 'higher level' physics concepts as quantum mechanics, relativity and symmetry breaking. Let us first look at the algebra.

| Time | Space | Mass | Charge |
|---|---|---|---|
| $i$ | **i j k** | 1 | *i j k* |
| pseudoscalar | vector | scalar | quaternion |

Each of the four fundamental parameters has its own algebra, and the whole system requires 8 basic units. Working out every possible combination requires 64.

|  |  |
|---|---|
| (±1, ± *i*) | 4 units |
| (±1, ± *i*) × (**i**, **j**, **k**) | 12 units |
| (±1, ± *i*) × (*i*, *j*, *k*) | 12 units |
| (±1, ± *i*) × (**i**, **j**, **k**) × (*i*, *j*, *k*) | 36 units |

This is a group of order 64 which requires only 5 generators. There are many ways of selecting these, but all such pentad sets have the same overall structure.

| Time | Space | Mass | Charge |
|---|---|---|---|
| *i* | **i j k** | 1 | *i j k* |

Take one of each of *i j k* on to each of the other three.

| | | |
|---|---|---|
| *i***k** | **i***i* **j***i* **k***i* | 1*j* |

Generators with this structure have exactly the same form as the 5 gamma matrices of the Dirac equation for the fermion, the fundamental equation of relativistic quantum mechanics.

Several consequences of this packaging are immediately apparent. The parameters time, space and mass acquire aspects of the parameter charge (quantization, conservation), while retaining their own respective characters as pseudoscalar, vector and scalar quantities.

| Energy | Momentum | Rest Mass |
|---|---|---|
| *i***k** | **i***i* **j***i* **k***i* | 1*j* |
| *i***k**E | *i***p** | *j*m |
| pseudoscalar | vector | scalar |

In effect, the packaging requires entirely new concepts to be created, as well as imposing conditions of quantization and relativistic connection. At the same time, the symmetry between the 'charge' components **k**, *i*, *j* is now broken, and they acquire characteristics which will ultimately determine the structure of the forces associated with them, while a fourth quantity (spin angular momentum) emerges as dual to the charge structure and ultimately carrying all the information that is contained in it (as required by the Noether's theorem extension discussed in the previous section).

| Weak charge | Strong charge | Electric charge |
|---|---|---|
| *i***k** | **i***i* **j***i* **k***i* | 1*j* |
| *i*w | **s** | e |
| pseudoscalar | vector | scalar |

It is not a coincidence that the fifth term here is the purely scalar / Coulombic source *m* or *e*. Interestingly, the *separate* Kaluza and Klein 5-D extensions of general relativity were aimed at the totally different objectives of explaining the origins of mass and electric charge. Clifford algebra (as used here in the multivariate 4-vector-quaternion combination) allows us to apply different concepts of 'dimensionality' to the same object, and we can see from the expression how an abstract concept of '10-dimensionality' could emerge, without needing an extension of the spatial dimensions beyond 3.

Now, because we have 'derived' energy-momentum-rest mass from space-time-mass, these are not truly independent or commutative systems, and, in quantum mechanics, they will anticommute (Heisenberg uncertainty). Only pure charge would be commutative with space-time-mass, but we cannot relate this to a direct observable. Putting together the components as a single package produces an interesting object:

$$(i k E + i \mathbf{p} + j m)$$

or, including all the possible sign variations

$$(\pm i k E \pm i \mathbf{p} + j m)$$

Now, for a single 'package' we have something like:

$$(\pm i k E \pm i \mathbf{p} + j m)(\pm i k E \pm i \mathbf{p} + j m) = E^2 - p^2 - m^2 = 0$$

The object is a *nilpotent*, or square root of zero.

Treated classically, this is simply Einstein's energy-momentum conservation equation (with $c = 1$). But we could also treat it as quantum mechanics by taking $E$ and $\mathbf{p}$ as operators and performing a canonical quantization on the left-hand bracket. Then, for a free particle, we would have something like:

$$(\pm i k \partial /\partial t \pm i \nabla + j m)(\pm i k E \pm i \mathbf{p} + j m) \exp^{(-i(Et - \mathbf{p}.\mathbf{r}))} = 0$$

giving us a nilpotent form of the Dirac equation. Ultimately, this is telling us that the complete range of space and time translations and rotations (or unlimited nonconservation) represented by the phase factor ($\exp(-i(Et - \mathbf{p}.\mathbf{r}))$) for the free particle is codified in the differential operator ($\pm i k \partial /\partial t \pm i \nabla + j m$).

The fact that the amplitude ($\pm i k E \pm i \mathbf{p} + j m$) ($\exp(-i(Et - \mathbf{p}.\mathbf{r}))$) resulting from the differentiation then squares to zero could then be seen as an expression of Pauli exclusion. However, Pauli exclusion is also true when the fermion is not free and the operators $E$ and $\mathbf{p}$ become something like

$$E = i \partial / \partial t \rightarrow i \partial / \partial t + e f + \ldots \text{ or other potentials}$$
$$\mathbf{p} = -i \nabla \rightarrow -i \nabla + e \mathbf{A} + \ldots \text{ or other potentials}$$

If we now assume that Pauli exclusion still requires a nilpotent amplitude, we will generate a particularly powerful version of quantum mechanics / quantum field theory, which requires a fermion to be defined only by the operator. The operator will then uniquely define a new phase factor (no longer a simple exponential) such that the resulting amplitude squares to zero.

$$\text{(operator acting on phase factor)}^2 = \text{amplitude}^2 = 0.$$

The operator now becomes a coding of the possible space and time variations which uniquely define a fermion, and which are decoded in the phase factor.

We can also now identify nilpotent mass ($jm$), nilpotent energy ($ikE$), spin angular momentum and nilpotent momentum ($i\mathbf{p}$) as having exactly the properties required for the dual group, and representing the respective duals to mass, time, charge and space. And this applies *only* to the nilpotent representation, because it is only in this structure that the mass dual is imaginary, as required. From an observational point of view, the second group of 4 are all that are needed, whereas the first group of 4 produce the entire ontology. However, though the two groups are dual, this duality is not absolute, and the Heisenberg uncertainty principle is the expression of this fact. The second group is not independent of the first, as it is constructed from it, and observation or epistemology is not independent of ontology, as ontology includes it. So the quantities, not being independent, do not commute. Classical conditions provide the closest approximation, creating a kind of overall 'phase space' structure.

**9 Interpretations of the nilpotent formalism**

The nilpotent formalism suggests many interesting interpretations in relation to quantum mechanics.
(1) ($\pm ikE \pm i\mathbf{p} + jm$) is defined only with respect to the entire quantum field; energy is conserved only with respect to the entire universe
(2) If each nilpotent state is necessarily unique, the formation of any new state, which is determined by the nature of all other nilpotent states, is a creation event within a unique birth-ordering.
(3) Locality is defined within the bracket ($\pm ikE \pm i\mathbf{p} + jm$), nonlocality outside it. The regions completely appropriate to each can be readily identified, and it becomes clear that, at the fundamental level, locality is meaningless without nonlocality, and *vice versa*. The group dualities are a required and absolute aspect of fundamental physics.
(4) If, as the group structure suggests, the total package adds up to a zero sum, then the 'rest of the universe' for a fermion (i.e. vacuum) is equivalent to what would be left after a fermion is created from nothing, or $-(\pm ikE \pm i\mathbf{p} + jm)$, meaning that no fermion has the same vacuum as any other.
(5) ($\pm ikE \pm i\mathbf{p} + jm$) is, in some sense, an angular momentum operator, whose three different components concern aspects of the three different things required to specify angular momentum (magnitude, direction and handedness), and which can also be directly related to the *e*, *s*, *w* interactions.

(6) Spin, helicity, *zitterbewegung*, bosons, baryons, partitioning of the vacuum, *CPT*, QED, QCD, QFD, renormalization, the mass gap, the Higgs mechanism, etc., are all exact and calculable consequences.[7]

(7) The idempotent objects *i*($\pm$ *ikE* $\pm$ *i***p** + *jm*), *j*($\pm$ *ikE* $\pm$ *i***p** + *jm*) and *k*($\pm$ *ikE* $\pm$ *i***p** + *jm*) appear to behave like vacuum 'reflections' (or *PCT* transformations) of the original fermionic state, responding to *s*, *e* and *w* interactions, and producing the additional terms in the Dirac spinor.

Essentially, the whole nilpotent approach produces a formalism that merges mainstream quantum mechanics and quantum field theory without needing second quantization, and providing additional results and interpretations. The nilpotent structure ($\pm$ *ikE* $\pm$ *i***p** + *jm*) only makes sense as a collective mode of the entire quantum field, and its intrinsically thermodynamic and dissipative nature (conserving energy only over the entire universe) makes the decoherence approach to quantum mechanics especially attractive.

Another thing that emerges is that the chirality intrinsic to the Dirac equation (filled weak vacuum) is retained, and not lost as it usually is in quantum field theory, where annihilation and creation, fermion and antifermion are treated as physically, as well as formally, symmetrical. This chirality is fundamental to nature (as a result of the 'packaging' mechanism) and does not need a 'cosmological' explanation. It is even present in (binary) number systems, where $-1$ has a different status to 1.

$$\ldots..1111111111111111111111111111 + 1 = 0$$

Taking this to its logical conclusion, we might hypothesize that, in the context of a filled vacuum, *gravity* (whose source is energy) might be taken as a nonlocal force, observable only through the local inertial reaction produced in discrete matter (so explaining its astonishing weakness and dissimilarity to the other interactions). In this context, the field equations of general relativity would take on an epistemological, rather than ontological, meaning, and something like 'dark energy' would be a natural result.[7] Such examples show that a more extensive view of physics, based on fundamental principles of symmetry, can generate interpretations which help to avoid some of the conventional dilemmas, as well as producing new formal results of interest.

**10 Conclusion**

Many people have claimed that quantum mechanics is 'strange' or contradictory to experience, or to other aspects of physics. On the contrary, if we take a more fundamental view of physics, based on the basic concepts needed to create its higher level structures, whether quantum or classical, we see that its 'strangeness' is a result of fundamental symmetries which are absolute, especially that between conservation and nonconservation, without which we could never have developed a successful description of the 'real' world.